\documentclass[usenatbib,useAMS]{mnras}
\usepackage{tablefootnote}
\usepackage{multirow}
\usepackage{xcolor}
\usepackage{hyperref}
\usepackage{graphicx}	
\usepackage{amsmath}	
\usepackage{amssymb}	
\usepackage{multicol}        
\usepackage{bm}		
\usepackage{url}
\usepackage[normalem]{ulem} 
\usepackage{caption}      
\usepackage{longtable}

\title[]{Multifrequency study of GPS sources with RATAN-600}

\author[Yu. Sotnikova et al.]{
Yu. Sotnikova,$^1$\thanks{E-mail: lacertae999@gmail.com} 
T.~Mufakharov,$^{1,2,3}$
M.~Mingaliev$^{1,2}$
and
A.~Mikhailov$^{1}$
\\
$^{1}$Special Astrophysical Observatory of RAS, Nizhny Arkhyz 369167, Russia\\
$^{2}$Kazan Federal University, 18 Kremlyovskaya St, Kazan 420008, Russia\\
$^{3}$Shanghai Astronomical Observatory, Chinese Academy of Sciences, Shanghai 200030, China\\
}

\date{6th CSS/GPS sources workshop proceedings, held in May 10-14, 2021 in Torun, Poland.}
\pubyear{2021}
\begin{document}
\label{firstpage}
\pagerange{\pageref{firstpage}--\pageref{lastpage}}
\maketitle

\begin{abstract}
We report the radio continuum properties for several samples of peaked spectrum (PS) radio sources. Broadband spectra of the objects were analysed using the RATAN-600 six-frequency (1.2--22~GHz) observations and available literature data, obtained on a time scale of 20--30~years.
We discuss statistical differences in radio properties for several AGN types with peaked spectra and focus on PS quasars at high redshifts ($z > 3$).
We confirm that a relatively small fraction (1--2\%) of bright PS sources can be considered as genuine GPSs when they have been monitored densely and for a long time. The contamination of GPS source samples by blazars is getting stronger as the redshift increases, and we confirm that it is underestimated due to lack of systematic multifrequency observations.
\end{abstract}


\section{Introduction}\label{sec1}
Gigahertz-Peaked Spectrum (GPS) sources are compact extragalactic radio sources with spectrum peaks around \mbox{$\sim$\,0.5--5~GHz}. GPS sources might represent either active galactic nuclei (AGNs) at the early stages of their evolution, or they could be radio galaxies frustrated by dense interstellar medium \citep{2021A&ARv..29....3O}.
The initial GPS source samples were mostly based on a combination of non-simultaneous radio measurements  \citep{1983A&A...123..107G,1985A&A...152...38S,1990A&AS...82..261O}, and possible variability could have influenced their GPS identification. In this paper, we will refer to GPS, compact steep-spectrum (CSS), megahertz-peaked spectrum (MPS), and high-frequency peaked (HFP) sources jointly as Peaked-Spectrum (PS) sources.

Among bright radio sources, a significant number are known to be of the PS type \citep{2009AN....330..128T,2011A&A...536A..15P,2013AstBu..68..262M}. They constitute more than 10\% of the bright radio sources but belong to different AGN types, including blazars during their flares. The main reason for the GPS source classification problem is the contamination by compact and beamed blazar-type sources \citep{2009AN....330..128T}. 

In Fig.~1 we present an example of the PS spectrum for a blazar and a classical GPS source. Their peaked spectra look similar, but have some differences: the blazar (J0646+44) has a broader peak and a flatter spectrum below and above peak frequency\footnote{We defined the spectral index from the power-law $S_{\nu}\sim\nu^{\alpha}$, where $S_\nu$ is a flux density at the frequency $\nu$, and $\alpha$ is a spectral index.}.
The variability index ${\rm Var}_{\rm radio}$ \citep{1992ApJ...399...16A} for J0646+44 reaches 0.65 at frequencies of 10--20 GHz on a long time scale (20--30 years), while for a classical GPS galaxy J1407+28 this value is only up to 0.16. 

In this paper we analyse broadband spectra of different PS source samples on a time scale of 20--30 years using the RATAN observations and available literature data. We have found spectral differences between GPS quasars (QSOs) and GPS galaxies (Gs), estimated the fraction of PS blazars, and discussed the peaked spectra of high-redshift quasars.

\begin{figure}
\centerline{\includegraphics[width=\columnwidth]{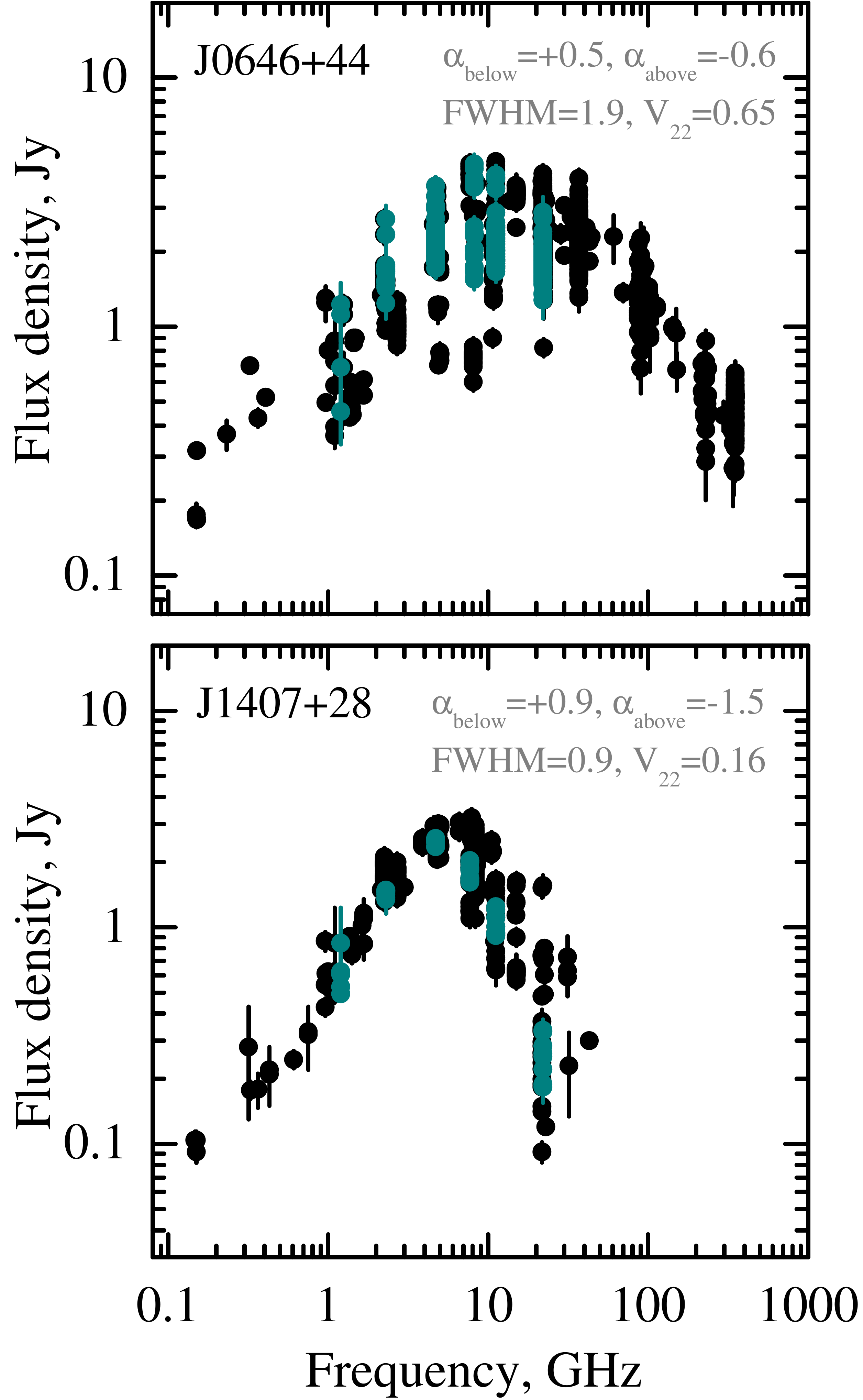}}
\caption{Radio continuum spectra for the PS blazar J0646+44 (top) and 
a classical GPS galaxy J1407+28 (bottom) compiled from the RATAN measurements (green) at 22, 11.2, 7.7/8.2, 4.7, 2.3 and 1.1 GHz and literature data (black). Both have a peaked spectrum shape during a long period of time. The well-known GPS galaxy J1407+28 has a very narrow (full width at half maximum ${\rm FWHM}=0.9$) spectral peak, steeper spectral indices above ($\alpha_{\rm above}$) and below ($\alpha_{\rm below}$) the maximum, and moderate flux density variability ${\rm Var}_{22}=0.16$.} 
\label{fig:f1}
\end{figure}

\section{GPS samples and RATAN observations in 2006--2021}\label{sec2}

We analysed the following GPS samples selected at different frequencies:

\textbf{Sample~1}, which is our target sample, consists of GPS galaxies and quasars
that were selected from the papers of
\cite{2001AJ....121.1306T,2005A&A...435..839T, 2007A&A...469..451T}. The total number of GPS sources and candidates is 122 (76 QSOs, 29 Gs, and 22 objects of an uncertain type). The observation period is 2006--2010
\citep{2012A&A...544A..25M}.

\textbf{Sample~2} is a complete sample of about 5000 radio sources from the CATS database \citep{1997BaltA...6..275V} with flux densities $S_{5\,{\rm GHz}} \geq 200$~mJy. The total number of PS sources in the sample is 467 (118 QSOs, 187 Gs, and 162 objects of an uncertain type), 8\% of them overlaps with Sample~1. The observation period is 2010--2019. The results are presented in \citet{2013AstBu..68..262M, 2019AstBu..74..348S}.

The comparison samples, \textbf{Sample~3} and \textbf{Sample~4}, are the complete samples of high-redshift quasars with flux densities $S_{1.4\,{\rm GHz}} \geq 100$~mJy at $z \geq 3$ and $S_{1.4\,{\rm GHz}} \geq 20$~mJy at \mbox{$z \geq 4$}. The total number of the objects is 102 and 37, respectively. 9\% of Sample~3 overlaps with Sample~1 and 18\% with Sample~2. 31\% of the Sample~4 also included in Sample~3. The observation period is 2017--2021 \citep{2021arXiv210914029S}.

The observations were carried out with the radio telescope RATAN-600 at 1.1, 2.3, 4.8, 7.7/8.2, 11.2, and 22 GHz. The advantage of the measurements is the possibility to obtain instantaneous radio spectra at several frequencies within \mbox{2--3}~minutes. We 
had been observing the samples at least 2--3 times per year and published the multifrequency catalogues of flux densities \citep{2012A&A...544A..25M,2013AstBu..68..262M,2019AstBu..74..348S}. The methods of observation and data reduction are described in \cite{2011AstBu..66..109T,2018AstBu..73..494T,2016AstBu..71..496U}. Broadband radio spectra were constructed and analysed together with additional data from the CATS database,\!\footnote{\url{www.sao.ru/cats/}} which contains a lot of radio catalogues and 
which we consider as a good GPS source searching tool.
The compiled radio data analysed in this paper cover a time period of 20--30 years.   

\section{Radio properties of GPS sources}\label{sec3}

We confirm that a relatively small fraction (1--2\%) 
of bright PS sources can be considered as genuine GPSs when they have been monitored densely and for a long time. For Sample~1, 48 out of 122 sources have a peaked spectral shape during 5 years and we classified them as genuine GPS \citep{2012A&A...544A..25M}, remaining sources have a flat, rising, or highly variable spectra (${\rm Var}_{\rm radio}>0.25$). For Sample~2, only 112 objects (2\%) can be considered as genuine GPS sources \citep{2013AstBu..68..262M}. 
In Sample~3, 46\% of sources of PS type, and 22\% of genuine GPS. There are 43\% of PS quasars in Sample~4, this study covers only several years of observations, and the quasars have not been studied long enough to discard the temporarily inverted spectra.

GPS candidates were selected assuming commonly accepted criteria for spectra of canonical GPS sources: spectral indices below and above the peak $\alpha_{\rm below}\geq+0.5$, $\alpha_{\rm above}\leq-0.8$, width of the spectra ${\rm FWHM}\sim1.2$ decades of frequency, and variability ${\rm Var}_{\rm radio}\leq0.25$ \citep{1997A&A...321..105D,1991ApJ...380...66O}.

We estimated statistical differences between GPS galaxies and quasars from Sample 1 and Sample 2 by comparing their average normalised radio spectra. The galaxies have a narrower spectral peak (${\rm FWHM}=1.4$) and steeper spectral indices ($\alpha_{\rm below}=+1.0$, $\alpha_{\rm above}=-0.8$) than those for quasars (${\rm FWHM}=1.8$, $\alpha_{\rm below}=+0.9$, and $\alpha_{\rm above}=-0.6$). We also revealed two PS groups with the narrowest spectra \citep{2012A&A...544A..25M,2013AstBu..68..262M,2019AstBu..74..348S}. The first group with ${\rm FWHM}<1$, located at $z < 1$, contains mostly galaxies. The second one with ${\rm FWHM}\sim 1.1$--$1.2$ at $z > 2$ contains mostly quasars.

\begin{figure*}
\centering
\begin{tabular}{lcr}
\begin{minipage}{0.33\linewidth}
\center{\includegraphics[width=\textwidth]{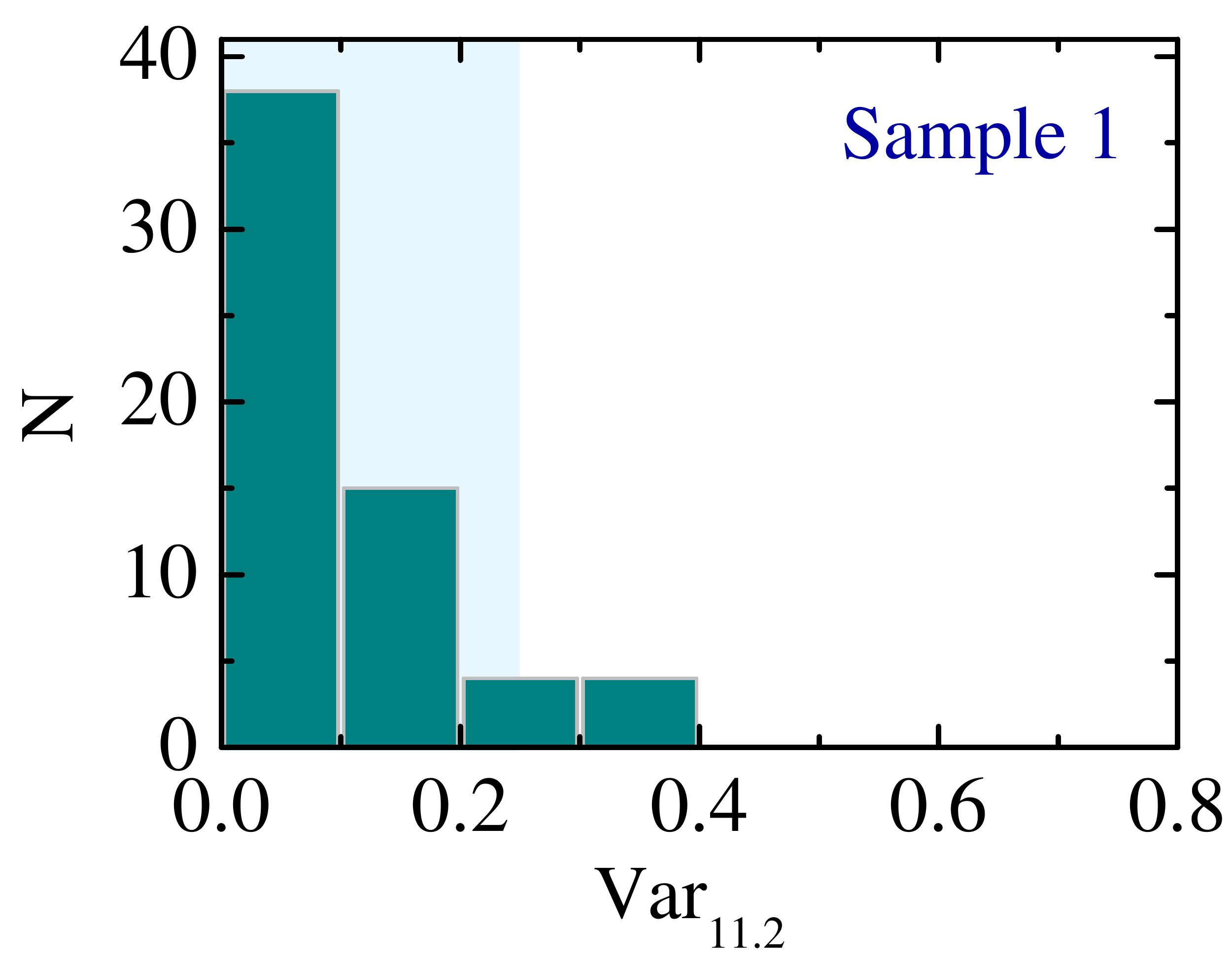}} 
\end{minipage}
\hfill
\begin{minipage}{0.33\linewidth}
\center{\includegraphics[width=\textwidth]{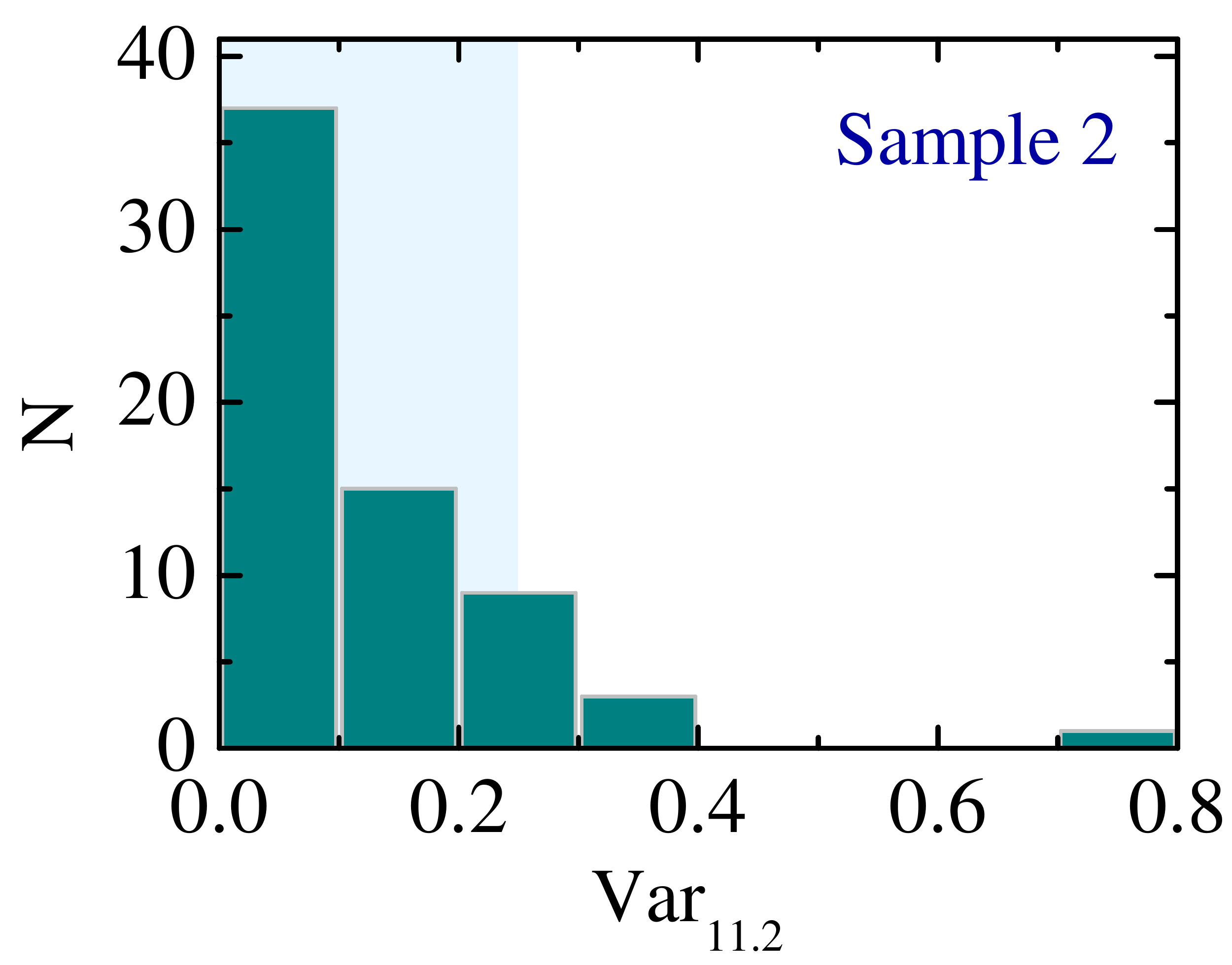}}  
\end{minipage}
\hfill
\begin{minipage}{0.33\linewidth}
\center{\includegraphics[width=\textwidth]{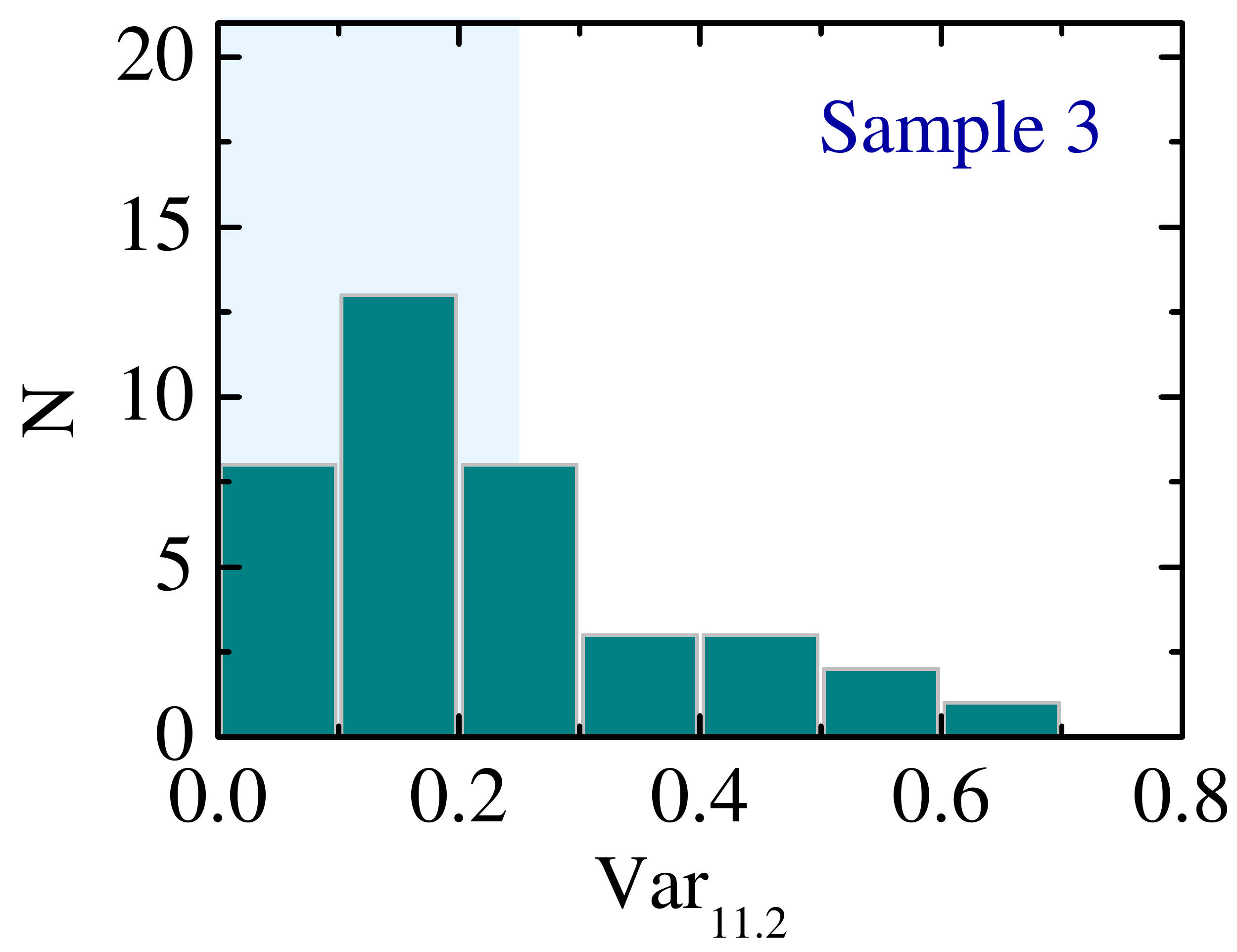}} 
\end{minipage}
\end{tabular}
\caption{The distribution of variability indices at 11.2 GHz for the three PS source samples. The blue area marks the accepted variability level for classical GPS sources.}
\label{fig:f2}
\end{figure*}

The radio variability is a key point for GPS classification. We compared average variability indices at 11.2 GHz (Fig.~\ref{fig:f2}) and obtained a value of about 0.05--0.08 for galaxies and 0.11--0.14 for quasars from Sample~1 and Sample~2 (Table~\ref{tab:var}). For the high-redshift quasars the average variability index ${\rm Var}_{11}=0.23$ and belongs to other distribution than that for PS sources from Samples 1 and 2 (at a significance level of 0.05 according to the Kolmogorov–Smirnov test).

\begin{table}[!]
\caption{Average variability index at 11.2 GHz for Samples 1, 2, and 3, estimated for time periods of 5, 12, and 4 years respectively. The standard deviations are given in parentheses.} 
\label{tab:var}
\centering
\begin{tabular}{cccc}
\hline
\multirow{2}{*}{Sample} & \multicolumn{2}{c}{${\rm Var}_{11.2}$} & Time period, \\
& G & QSO & years \\ 
\hline
\hline
1 & 0.05 (0.04) & 0.11 (0.10) & 5 \\ 
2 & 0.08 (0.06) & 0.14 (0.13) & 12 \\ 
3 & -    & 0.23 (0.15) & 4 \\ 
\hline
\hline
\end{tabular}
\end{table} 

\subsection{PS Sources at High Redshifts}

PS quasars are often found at high redshifts \citep{1990MNRAS.245P..20O,1998A&AS..131..303S,2015MNRAS.450.1477C}.
We observed two comparison samples of quasars at $z \geq 3$ ($S_{1.4\,{\rm GHz}}~\geq$~100 mJy) and $z \geq 4$ ($S_{1.4\,{\rm GHz}}~\geq$~20 mJy) in 2017--2021 and revealed that the peaked spectrum is a common feature for them, almost half of the objects had the observed maximum in their radio spectra at the GHz frequencies \citep{2021arXiv210914029S}. Using the RATAN measurements \citep{2021MNRAS.503.4662M}, we defined the newly discovered most distant blazar PSO J047.4478+27.2992 at $z = 6.1$ \cite{2020A&A...635L...7B} as an MPS candidate. Comparison with the other distant blazars at $z > 5$ (J1026+25, J0906+69, and J1648+46) revealed the PS for three out of four of them. One explanation is the predominant contribution of the bright compact core emission in quasars which dominate the high-redshift source population due to the Doppler boosting effect.

\subsection{PS Blazars}
We analysed the radio spectra of blazars from the Roma-BZCAT catalogue by \citet{2009A&A...495..691M,2015Ap&SS.357...75M}. The BZCAT catalogue contains 3561 blazars, and their classification is based on spectral energy distribution features, namely strong non-thermal emission from radio waves to $\gamma$-rays and the evidence of relativistic beaming.
We constructed their radio spectra using the published radio measurements from the Astrophysical CATalogs support System (CATS) database and have revealed 504 (14\%) sources with spectral peak. 
For 22\% of them there are only sparse data (up to 20 measurements) from few observing epochs, and long-term multifrequency observations are needed to classify their radio spectra reliably. We suppose that the blazar contamination is underestimated due to the lack of long-term multifrequency observations.

We have compiled the BLcat catalogue\footnote{https://www.sao.ru/blcat/} \citep{2014A&A...572A..59M} of blazars using the RATAN-600 multifrequency measurements in 2004--2021 and radio data available from the literature. The catalogue contains 1598 blazars of different types and allows studying their radio continuum spectra evolution on a long time scale at the frequencies of 1.1--22 GHz. 
We analysed the radio spectra of the BLcat objects and found 205 blazars that have peaked spectrum shape in at least one epoch. 42 of them (20\%) have peaked spectrum for a long time, which is about 2.6\% of the BLcat blazars. We consider the BLcat as an effective GPS source searching tool which allows jet-beamed objects to be excluded from GPS source samples.

\section{Summary}\label{sec4}

We have analysed radio properties of several different PS source samples and concluded that a reliable spectral classification is still in demand and multifrequency monitoring is the best tool for GPS identification. According to multifrequency and long-term observations a very small fraction (1--2\%) of bright PS radio sources can be considered as genuine GPS sources. The radio spectra of PS blazars and genuine GPS sources differ statistically. At the high redshifts ($z > 3$), about half of bright radio sources have peaked spectra. We revealed peaked spectra in about 14\% of the blazars from the Roma-BZCAT catalogue, but more than 20\% of BZCAT blazars do not have long-term radio measurements for reliable spectral classification. The contamination of GPS samples by blazars is getting stronger with the redshift increasing.

\section*{Acknowledgments}

The observations were carried out with the RATAN-600 scientific facility, which is supported by the Ministry of Science and Higher Education of the Russian Federation. The study has been supported by the Kazan Federal University Strategic Academic Leadership Program.

\bibliographystyle{mnras} %
\bibliography{Wiley-ASNA}%



\bsp	
\label{lastpage}
\end{document}